\documentclass{elsart}

\usepackage{supertabular}

\def\astrobj#1{#1}
\def\bibcode#1{}
\def\url#1{\texttt{#1}}
\def\asec{\ifmmode ^{\prime\prime}\else$^{\prime\prime}$\fi}

\begin{document}

\runauthor{Greiner}

\begin{frontmatter}
\title{Catalog of supersoft X-ray sources}

\author[]{Jochen Greiner}

\address[]{Astrophysical Institute Potsdam, An der Sternwarte 16, 14482
  Potsdam, Germany}

%\thanks[Someone]{Partially supported by ....}

\begin{abstract}
This catalog comprises an up-to-date (December 1999) list of luminous
($>$10$^{36}$
erg/s), binary supersoft X-ray sources. This electronic version (including the
accompannying Web-pages) supersedes the printed version of Greiner (1996).
\end{abstract}
\begin{keyword}
supersoft X-ray sources -- binaries -- hydrogen burning \\
95.80.+p       Catalogues, atlases, databases;
97.80.Gm       Cataclysmic binaries;
97.80.Jp       X-ray binaries;
98.62.Mw       Infall, accretion, and accretion disks;
\end{keyword}
\end{frontmatter}

%\section{Aim}
%This catalog comprises an up-to-date (December 1999) list of luminous 
%($>$10$^{36}$
%erg/s), binary supersoft X-ray sources. This Web-version supersedes the
%printed version of Greiner (1996).

\section{Introduction}

After the discovery of supersoft X-ray sources with Einstein Observatory
observations, the ROSAT satellite with its PSPC detector has discovered
about four dozen new supersoft sources and has thus established luminous
supersoft X-ray sources (SSS) as a new class of objects. Though many
different classes of objects emit supersoft X-ray radiation (defined here as
emission dominantly below 0.5 keV which corresponds to effective
temperatures of the emitting objects of $<$50 eV), we consider here sources
with bolometric luminosities in the range $10^{36}-10^{38}$ erg/s. Optical
observations have revealed the binary nature of several of these objects. A
white dwarf (WD) model, the so-called close-binary supersoft source (CBSS)
model, is perhaps the most promising (van den Heuvel et al. 1992; Rappaport,
DiStefano, Smith 1994, Kahabka \& van den Heuvel 1997). 
It invokes steady-nuclear burning on the surface of
an accreting WD as the generator of these systems' prodigious flux. Indeed,
SSS temperatures and luminosities as derived from the X-ray data suggest an
effective radius comparable to that of WDs. Eight SSSs have orbital periods
between approximately 4 hrs and 3.5 days. These are the candidates for the
CBSS model. Mass transfer rates derived from the CBSS model are in the right
range for steady nuclear burning of the accreted matter.

This catalog comprises an up-to-date (December 1999) list of luminous 
($>10^{36}$
erg/s) supersoft X-ray sources. We include in this catalog accreting binary
sources of high luminosity which are thought to be in a state of (steady or
recurrent) hydrogen burning. Since CAL 83, the prototype, is known to have
an ionisation nebula (Pakull and Motch 1989), and further supersoft binaries
are expected to also have one, we include also sources associated with very
luminous planetary nebulae. Not included are the low-luminosity objects like
single (i.e. non-interacting) white dwarfs and magnetic cataclysmic
variables, and PG 1159 stars which reach similar luminosities but form a
rather distinct class (e.g. Dreizler et al. 1995). Excluded are also
supersoft active galactic nuclei which reach luminosities up to 
10$^{45}$ erg/s,
and the recently found examples of large-amplitude outbursts of supersoft
X-ray emission which have been interpreted as tidal disruption events (e.g.
Komossa \& Greiner 1999).
Since most of the new sources are X-ray discoveries, the final inclusion in
the group of luminous close binary supersoft sources has to await the
optical identification. Only then a distinction is possible among the
various and quite different types of objects which show a supersoft X-ray
spectrum (i.e. emission only below 0.5 keV) but have different luminosities.
Due to this fact of necessary follow-up optical observations, it can well
happen that a source is included in an early version of the catalog but
later turns out to be of a different type. An example is 
\astrobj{RX J0122.9--7521}
which has long been thought to be a SMC supersoft source (Kahabka et al.
1994), but has been identified as a galactic PG 1159 star (Cowley et al.
1995, Werner et al. 1996), and therefore has been removed from this catalog.

\section{Einstein, ROSAT and beyond...}

The two most famous supersoft X-ray sources, \astrobj{CAL 83} and 
\astrobj{CAL 87} (Long et al.
1981), have been discovered with Einstein satellite observations. ROSAT
observations established these sources as a distinct class in the early
nineties, and the majority of the X-ray measurements have been performed
with the ROSAT position sensitive proportional counter (PSPC) during
1990-1995, yielding a source position accuracy of about 25\asec. The PSPC with
its spectral resolution of about 50\% below 1 keV has been used in nearly all
cases to discover the supersoft X-ray spectrum. During the years 1995-1999
the high-resolution imager (HRI) on ROSAT has been used to improve the
coordinates of the newly detected sources down to typically 10\asec\ and to
monitor the long-term X-ray intensity. At these soft energies, the HRI count
rates are typically a factor of 7.5--8 smaller than those of the PSPC (David
et al. 1994, Greiner et al. 1996a). Since 1997, some of the brightest
supersoft X-ray sources have been also observed with the low-energy
concentrator spectrometer (LECS) onboard BeppoSAX. As of the time of this
writing, there is only one SSS (U Sco) which has not been observed by ROSAT,
and consequently has no entry in the "ROSAT count rate" field.

With the Chandra and XMM missions starting regular observations, a wealth of
new information on the X-ray properties of these supersoft sources can be
expected, as well as new discoveries. In particular, the better energy
resolution, through-put and location accuracy will improve our
understanding. Therefore, this will most probably not be the last version of
a SSS catalog.

\section{Organisation of the catalog}

The catalog is organised as follows: The catalog consist of four major
parts: a master table, a bibliography, a query form, and the individual
source pages. While the master table should provide some basic numbers, the
individual source pages are thought to provide some more parameters as well
as the links to the details behind these numbers like graphs and references.
\begin{itemize}
\item The master table gives an overview of all sources with their main
     characteristics, including ROSAT PSPC count rate, temperature,
     luminosity, type of binary and orbital period (if known).

\item The bibliography contains relevant papers, sorted chronologically, and
     alphabetically within each year. It is supposed to be complete for the
     CBSS type sources, while for the other sources only those papers where
     included which provided input for the source parameters. An attempt has
     been made to provide a direct link to all papers which are
     electronically available (either directly from the journal pages or
     through ADS).

\item The query form allows you to interactively create new tables for a
     sub-sample of sources and arbitrary parameters of your choice.

\item The individual source tables provide information on a variety of source
     parameters, including figures (e.g. of spectra or light curves) and
     links to references. All source related numbers appear in red, and
     links in blue. Blue "numbers" correspond to the reference with the same
     number at the bottom of the page, and the underlying link points to the
     bibliography. Blue "words" contain links to figures which will open in
     a separate window (this separate window is used for all subsequent
     links too). The data are organized as follows:
\begin{itemize}
  \item The top part contains the name, coordinates (equinox 2000.0), the
          type of binary, the ROSAT count rate, a link to the discovery
          paper and a finding chart. The "Discovery" reference refers to the
          first paper which realised the luminous supersoft X-ray emission.
          Some of the sources have been known already for decades at this
          time, so there may be many references listed which appeared
          earlier than the "Discovery" reference. If a source is not
          optically identified, the R.A./Dec. numbers are the X-ray
          positions with uncertainties as given in the above section of
          ROSAT related issues. Otherwise, the optical positions are given
          with typical uncertainties of 1''. All coordinates are equinox
          2000.0.
    \item The next (table) block contains general data of the system like
          distance, orbital period, brightness and color. The radial
          velocity amplitude is given for the HeII 4686 A line; note that it
          is common in CBSS that other emission lines as well as the Balmer
          absorption lines have quite different velocity amplitudes! The
          right column provides links to spectra in all wavelength ranges.
          For sources supposed to belong to an external galaxy, the galaxy
          name is given instead of the presently known distance. This is
          motivated by recent changes in the distance determination of the
          LMC. This in turn also affects the distance to M31 because the
          distance ratio of LMC and M31 is more accurately known than the
          corresponding absolute distances. Note that different authors may
          have used different distances for sources within the same galaxy,
          so that a simple comparison of the luminosities may be misleading!
          The galactic absorbing column (NH$_{gal}$) is taken from Dickey and
          Lockman 1990) and is given for comparison with the values derived
          from the X-ray fits.
    \item The next table block provides details on the X-ray spectral
          fitting. Here, a distinction is made between the use of a simple
          blackbody model versus more sophisticated white dwarf atmosphere
          models. While in general the blackbody overestimates the
          bolometric luminosity by a large factor (10--100), it should be
          noted that there are a variety of white dwarf atmosphere models on
          the market which have been used by different authors for different
          sources. Thus, special care is needed when comparing these
          parameters among different sources!
    \item The next table block gives hints on the variability of the source,
          separated into orbital modulation and non-orbital, intrinsic
          variability. The comments are short, so please use the links to
          check out the original papers to get a complete picture about the
          source's behaviour.
    \item The final table block applies to only a few sources, for which
          optical measurements of an ionisation nebula or bipolar outflow or
          radio measurements have been conducted. The radio fluxes (limits)
          are given for 3.5 cm.
    \item Finally, the references are given in full with all co-authors, and
          stating the important pieces of new information (this reflects
          purely my subjective view, and you should contact me if something
          is missing or wrong.) The references are sorted in time of
          appearance, so that the numbers in the data blocks will not change
          when the catalog is updated.
   \end{itemize}
\end{itemize}

\section{Warning and Request}

I would like to emphasise that every user of this catalog should spare no
pains to consult the original papers in order to avoid propagation of my
errors in the literature. I will keep this catalog updated, and would
appreciate (1) being informed on any errors/omissions users might discover
and (2) getting reprints of papers on supersoft sources to be included in
the next version.

\section{Availability}

%This catalogue is freely available for everybody, but should not be copied
%or sold to other parties. It is an unfortunate aspect of modern science
%funding that impersonal and statistical measures are used to assess the
%productivity and usefulness of persons/programs. Therefore, if the catalogue
%was helpful for your research work, the following reference would be
%appreciated: "This research has made use of the electronic catalog of
%supersoft X-ray sources available at URL
%http://www.aip.de/~jcg/sss/ssscat.html and maintained by J. Greiner".

The full Web-based catalog is available at URL \\
\url{http:/$\!$/www.aip.de/People/JGreiner/sss/ssscat.html}.

\bigskip
{\it Acknowledgements:}
I appreciate the help of many collegues who sent preprints and reprints of
their work. Special thanks to Rosanne Di Stefano for her steady
encouragement to produce this catalog and for extensive discussion on its
content. The accompanying Web-version would not have been possible without 
the enormous help by Arne Rau and Robert Schwarz (both AIP) who wrote 
the Perl and Java scripts for
the generation of the source pages. I apologise to anyone whose paper
slipped through the literature search. The ROSAT project was supported 
by the German Bundesministerium f\"ur Bildung, Forschung, Wissenschaft und
Technik (BMBW/DARA) and the Max-Planck-Society. This research has made use
of the Simbad database, operated at CDS, Strasbourg, France.

\newpage

\small
\begin{table} 
   \caption{Summary of all known supersoft X-ray sources with
              luminosities above 10$^{36}$ erg/s excluding PG 1159-type stars
              and supersoft AGN.
              Given are for each source the name (column 1),
              the ROSAT PSPC count rate (2)
              the best fit X-ray temperature with bb indicating blackbody
              and wd white dwarf atmosphere models (3),
              the bolometric luminosity (4),
              the type of system (5),
              and the binary period (6).
     }
\end{table}
\tablehead{\hline\noalign{\smallskip}
  \hspace{10mm} Name \hspace{12mm} & Countrate$^{(1)}$ & \hspace{6mm} 
              T$^{(2)}$ \hspace{6mm} &
  \hspace{7mm} L$_{\rm bol}$ \hspace{7mm} &   Type$^{(3)}$ & ~Period \\
               & (cts/s)  & (eV)   & (erg/s) &        & (days) \\
\noalign{\smallskip}\hline\noalign{\smallskip}}
\tabletail{\hline\noalign{\smallskip}}
\begin{supertabular}{lccccc}
%\begin{tabular}[h]{lccccc} 
%\hline 
\noalign{\medskip}
\multicolumn{6}{c}{\bf Galactic Sources} \\ 
\noalign{\medskip}
\astrobj{RX J0019.9+2156}  & 2.0 & $25-37$ (wd) & $(3-9)\cdot10^{37}$ & CBSS & $1.0-1.35$ \\ 
\astrobj{RX J0925.7-4758} & 1.0 & $70-75$ (wd) & $(3-7)\cdot10^{35}$ & CBSS & $3.55-4.03$ \\ 
\astrobj{GQ Mus} & 0.1 & $25-35$ (bb) & $(1-2)\cdot10^{38}$ & N & 0.0588 \\ 
1E 1339.8+2837 & $0.01-1.1$ & $20-45$ (bb) & $1.2\cdot10^{34}-1.2\cdot10^{36}$ &  &  \\ 
\astrobj{AG Dra} & $0.01-1.0$ & $10-15$ (bb) & $9.5\cdot10^{36}$ & Sy & $549-554$ \\ 
\astrobj{U Scorpii} &  & $74-76$ (wd) & $(8-60)\cdot10^{36}$ & RN & 1.2306 \\ 
\astrobj{RR Tel} & 0.18 & 12 (wd) & $1.3\cdot10^{37}$ & Sy & \\ 
\astrobj{V Sge} & $0.001-0.02$ & $<80 (bb)$ & $(1-10)\cdot10^{37}$ & CV  & 0.514195 \\ 
\astrobj{V1974 Cyg} & $0.03-76$ & $30-51$ (wd) & $2\cdot10^{38}$ & N & 0.0812\\
\astrobj{V751 Cyg} & $0.039-0.11$ & 15 (bb) & $6.5\cdot10^{36}$ & VY Scl &  \\
\noalign{\medskip}
\multicolumn{6}{c}{\bf Large Magellanic Cloud} \\ 
\noalign{\medskip}
\astrobj{RX J0439.8-6809} & 1.35 & $20-25$ (wd) & $(10-14)\cdot10^{37}$ & CBSS & 0.1404 \\ 
\astrobj{RX J0513.9-6591} & $<0.06-0.2$ & $30-40$ (bb) & $(0.1-6)\cdot10^{37}$ & CBSS & 0.76278 \\ 
\astrobj{Nova LMC 1995} & 0.061 & $20-40$ (wd) & $1.5\cdot10^{38}$ & N &  \\ 
\astrobj{RX J0527.8-6954} & $0.004-0.25$ & $18-45$ (bb) & $(1-10)\cdot10^{37}$ & CBSS & 0.3926 \\ 
\astrobj{RX J0537.7-7034} & $<0.002-0.02$ & $18-70$ (bb) & $(0.6-2)\cdot10^{37}$ & CBSS & 0.125 \\ 
\astrobj{CAL 83} & $<0.035-0.98$ & $39-60$ (wd) & $<2\cdot10^{37}$ & CBSS & 1.04 \\ 
\astrobj{CAL 87} & 0.09 & $63-84$ (wd) & $(6-20)\cdot10^{37}$ & CBSS & 0.44267 \\ 
\astrobj{RX J0550-7151} & $<0.02-0.9$ & $25-40$ (bb) &  &  &  \\ 
\noalign{\medskip}
\multicolumn{6}{c}{\bf Small Magellanic Cloud} \\ 
\noalign{\medskip}
\astrobj{1E 0035.4-7230} & 0.33 & $40-50$ (wd) & $(0.8-2)\cdot10^{37}$ & CBSS & 0.1719 \\
\astrobj{RX J0048.4-7332} & $0.19-0.33$ & $25-45$ (wd) & $(1-8)\cdot10^{38}$ & Sy &  \\ 
\astrobj{1E 0056.8-7154} & 0.29 & $30-40$ (wd) & $2\cdot10^{37}$ & PN &  \\ 
\astrobj{RX J0058.6-7146} & $<0.001-0.7$ & $15-70$ (bb) & $2\cdot10^{36}$ &  &  \\ 
\noalign{\medskip}
\multicolumn{6}{c}{\bf Andromeda Galaxy (M\,31)} \\ 
\noalign{\medskip}
RX J0037.4+4014  & $0.8\cdot10^{-3}$ & $26-37$ (wd) &  &  &  \\ 
RX J0037.4+4015 & $0.3\cdot10^{-3}$ & $<34$ (wd) &  &  &  \\ 
RX J0038.6+4020 & $1.7\cdot10^{-3}$ & $28-38$ (wd) &  &  &  \\ 
RX J0039.4+4050 & $2.9\cdot10^{-3}$ & $49$ (wd) &  &  &  \\ 
RX J0039.6+4054 & $0.4\cdot10^{-3}$ & $<36$ (wd) &  &  &  \\ 
RX J0039.7+4030 & $2.0\cdot10^{-3}$ & $<38$ (wd) &  &  &  \\ 
RX J0040.0+4100 & $2.0\cdot10^{-3}$ & $52-58$ (wd) &  &  &  \\ 
RX J0040.1+4021 & $0.5\cdot10^{-3}$ & $>51$ (wd) &  &  &  \\ 
RX J0040.4+4004 & $0.8\cdot10^{-3}$ & $28-36$ (wd) &  &  &  \\ 
RX J0040.7+4015 & $1.3\cdot10^{-3}$ & $28-38$ (wd) &  &  &  \\ 
RX J0041.5+4040 & $0.3\cdot10^{-3}$ & $31-34$ (wd) &  &  &  \\ 
RX J0041.8+4015 & $3.2\cdot10^{-3}$ & $45$ (wd) &  &  &  \\ 
RX J0041.8+4059 & $0.5\cdot10^{-3}$ & $29-35$ (wd) &  &  &  \\ 
RX J0042.4+4044 & $1.7\cdot10^{-3}$ & $28-38$ (wd) &  &  &  \\ 
RX J0042.4+4048 & $0.6\cdot10^{-3}$ & $39-62$ (wd) &  &  &  \\ 
RX J0042.6+4043 & $1.6\cdot10^{-3}$ & $>55$ (wd) &  &  &  \\ 
RX J0042.6+4159 & $1.8\cdot10^{-3}$ & $>63$ (wd) &  &  &  \\ 
RX J0042.8+4115 & $40.1\cdot10^{-3}$ & $65$ (wd) &  &  &  \\ 
RX J0043.3+4120 & $6.7\cdot10^{-3}$ & $59-62$ (wd) &  &  &  \\ 
RX J0043.5+4207 & $2.2\cdot10^{-3}$ & $32-39$ (wd) &  &  &  \\ 
RX J0043.7+4127 & $1.2\cdot10^{-3}$ & $46-60$ (wd) &  &  &  \\ 
RX J0043.9+4151 & $0.9\cdot10^{-3}$ & $43$ (wd) &  &  &  \\ 
RX J0044.0+4118 & $2.5\cdot10^{-3}$ & $32-39$ (wd) &  &  &  \\ 
RX J0044.2+4026 & $7\cdot10^{-5}$ & $>69$ (wd) &  &  &  \\ 
RX J0044.4+4200 & $1.2\cdot10^{-3}$ & $47-56$ (wd) &  &  &  \\ 
RX J0045.4+4154 & $<10^{-5}-0.03$ & $72-73$ (wd) & $(5-10)\cdot10^{37}$ & & \\
RX J0045.4+4219 & $1.2\cdot10^{-3}$ & $51-59$ (wd) &  &  &  \\ 
RX J0045.5+4206 & $3.1\cdot10^{-3}$ & $35-40$ (wd) & $7\cdot10^{37}$ &  &  \\ 
RX J0046.1+4136 & $0.3\cdot10^{-3}$ & $>51$ (wd) &  &  &  \\ 
RX J0046.2+4138 & $1.1\cdot10^{-3}$ & $29-38$ (wd) &  &  &  \\ 
RX J0046.2+4144 & $2.1\cdot10^{-3}$ & $27-38$ (wd) &  &  &  \\ 
RX J0046.3+4238 & $3.1\cdot10^{-3}$ & $53-62$ (wd) &  &  &  \\ 
RX J0047.6+4132 & $0.3\cdot10^{-3}$ & $>60$ (wd) &  &  &  \\ 
RX J0047.6+4205 & $0.3\cdot10^{-3}$ & $<34$ (wd) &  &  &  \\ 
\noalign{\medskip}
\multicolumn{6}{c}{\bf NGC 55} \\ 
\noalign{\medskip}
\astrobj{RX J0016.0-3914}  & 0.0045 & $23 \pm 30$ (bb) & $(9)\cdot10^{37}$ &  &  \\ 
\noalign{\smallskip}
\hline 
\end{supertabular}

    \noindent{\small $^{(1)}$\,Countrates in the ROSAT PSPC corrected
                for vignetting, i.e. absorbed on-axis count rates. Count rates
                in the HRI have been converted to PSPC rates
                using a conversion factor of PSPC/HRI = 7.8
                (Greiner \etal\ 1996a).\\
              $^{(2)}$\,Temperatures for the M31\,sources are the maximum
                 blackbody temperatures derived from the hardness ratios
                 at the appropriate absorbing column (Greiner\,\etal\,1996b).\\
               $^{(3)}$ CBSS = close-binary supersoft X-ray source, N = nova,
                 Sy = symbiotic binary, RN = recurrent nova, CV = cataclysmic 
                 variable, PN = planetary nebula.
                }

\end{document}